\documentstyle[12pt,graphicx]{article}
\newcommand{\D}{{\not\hspace{-0.8ex}\pi}}
\begin{document}
\title{Effects of Anomalous Magnetic Moment in the Quantum Motion of 
Neutral Particle in Magnetic and Electric Fields Produced by a Linear 
Source in a Conical Spacetime}
\author{E. R. Bezerra de Mello \thanks{E-mail: emello@fisica.ufpb.br}\\
Departamento de F\'{\i}sica-CCEN\\
Universidade Federal da Para\'{\i}ba\\
58.059-970, J. Pessoa, PB\\
C. Postal 5.008\\
Brazil}
\maketitle       
\begin{abstract} 
In this paper we analyse the effect of the anomalous magnetic moment on the 
non-relativistic quantum motion of a neutral particle in magnetic and electric
fields produced by linear sources of constant current and charge density, 
respectively.
\vspace{1pc}
\end{abstract}

\maketitle

\section{Introduction}
The role played by topology in the physical properties of a variety of
system has been analysed in different areas as gravitation theory and
condensed matter physics, for example. Topological defects appears in 
classical filed theory, as consequence of spontaneously broken of
gauge symmetry as monopoles, strings and  wall \cite{Vilenkin}. In condensed 
matter physics they are vortices in superconductors or superfluid 
and domain wall in magnetic materials \cite{Kleiner}, and dislocations or
disclinations in disordered solid or liquid crystal \cite{Kleman}.
Cosmic strings and disclinations are linear defects that changes the topology
of the the medium in a similar way. In fact the similarity between these two
objects goes beyond the  topology: for some applications both kind of defects
can be treated by the same geometrical methods \cite{Katanaev}.

Although the spacetime produced by a thin, infinite, straight cosmic string 
is locally flat \cite{Vilenkin1}, globally it is not. In fact the lack of
global flatness is responsible for the existence of an induced electrostatic
self-interaction on a point charged particle \cite{Linet,Smith}, and also an
induced electrostatic and magnetostatic self-interaction on a long straight
wire which present constant linear density of charge and current parallel to
the topological defect \cite{Mello}, gravitational analog of the Aharonov-Bohm 
effect \cite{Dowker}, among others \cite{Frolov}.

The metric tensor associated with infinite straight cosmic string lying along
the $z-$axis can be given in cylindrical coordinate system by the following line 
element:
\begin{equation}
\label{CS}
d^2s=-c^2dt^2+dr^2+\alpha^2r^2d\phi^2+dz^2 \ ,
\end{equation}
where $r\geq 0$ and $0\leq \phi \leq 2\pi$. The parameter $\alpha$ is given
in terms of the linear mass density of the cosmic string $\mu$ by $\alpha=1
-4\mu/c^2$, which is smaller than unity. The linear defect in crystal liquid
disclination, is obtained by either removing or inserting a wedge of material.
Being $\sigma$ the angle that defines the wedge, the metric of the 
disclinated medium is, in cylinder coordinates, similar to (\ref{CS}) according 
to the geometric approach given in  Ref. \cite{Katanaev}, however $\alpha=
1+\sigma/2\pi$. The space section corresponding to a conical one and the only
nonzero components of the Riemann curvature tensor and Ricci tensor are give
by:
\begin{equation}
R^{12}_{12}=R^1_1=R^2_2=2\pi\frac{1-\alpha}{\alpha} \ \delta^2(\vec{r}) \ ,
\end{equation}
where $\delta^2(\vec{r})$ is the two dimensional delta function in flat
space \cite{Sokolov}. From the above expression follows that if $0<\alpha
<1$, the defect presents a positive curvature and a planar deficit
angle. On the other hand for disclinated medium the $\alpha$ parameter can
be greater than unity for positive $\sigma$, and in this case the defect
carries negative curvature and excess of planar angle, which corresponds to
an anticonical spacetime.

As it was shown in \cite{Mello} the presence of linear electric or magnetic
source in the spacetime of a cosmic string parallel to the latter, become
subject to a induced self-forces per unit length given by
\begin{equation}
\frac{\vec{F}_{Ele.}}l=\frac{(p-1)\lambda^2}r\hat{r} 
\end{equation}
and
\begin{equation}
\frac{\vec{F}_{Mag.}}l=-\frac{(p-1)}r\frac{j_0^2}{c^2}\hat{r} \ ,
\end{equation}
where $p=1/\alpha$, $\lambda$ the linear electric charge density and $j_0$ the 
current. We can see that for $p$ bigger than unity the electric self-force
becomes repulsive and the magnetic one attractive. For $p$ smaller than unity 
the opposite situation occur.

The electric and magnetic fields produced by these two linear sources can be
obtained as shown below:
\\
$(a)$ {\bf Electric Field}\\
Let us first calculate the scalar potential produced by the uniform linear
charge density. In general the scalar potential is given by the Poison equation
\begin{equation}
\nabla^2\Phi(\vec{r})=-4\pi\rho(\vec{r}) \ ,
\end{equation}
where in this case $\rho(\vec{r})=\lambda\delta^2(\vec{r}-\vec{r}_0)$. 
Calculating the solution to the Green function in the section $z=\ t=\ const.$
\begin{equation}
\nabla^2G^2_p(\vec{r},\vec{r}_0)=-4\pi\delta^2(\vec{r}-\vec{r}_0) \ ,
\end{equation}
one finds
\begin{equation}
G_p^2(\vec{r},\vec{r}_0)=-\ln\left[r^{2p}+r_0^{2p}-2(rr_0)^p\cos(\phi-\phi_0)
\right] \ .
\end{equation}
So the scalar potential reads
\begin{equation}
\label{phi} 
\Phi(\vec{r})=\lambda \ G_p^2(\vec{r},\vec{r}_0) \ .
\end{equation}
The electric field is obtained taking its negative gradient: $\vec{E}=-
\vec{\nabla}\Phi(\vec{r})$.
\\  
$(b)$ {\bf Magnetic Field}\\
The vector potential associated with this linear current can be calculated by
\begin{equation}
\vec{A}(\vec{r})=\frac1c\int \ d^2r \ G_p^2(\vec{r},\vec{r'})
\ \vec{J}(\vec{r}) \ ,
\end{equation}
with $\vec{J}(\vec{r})=j_0\hat{z}\delta^2(\vec{r'}-\vec{r}_0)$. Substituting
this expression inside the integral above we obtain:
\begin{equation}
\label{A}
\vec{A}(\vec{r})=\frac{j_0}c\hat{z} \ G_p^2(\vec{r},\vec{r}_0) \ .
\end{equation}
Finally the magnetic field can now be calculated by taking its rotational: 
$\vec{B}=\vec{\nabla}\times\vec{A}$.

The propose of this paper is to analyse the non-relativistic quantum motion of
an uncharged spin-$1/2$ particle with an anomalous moment interacting with
the electric and magnetic fields produced by a linear electric and magnetic
sources in the cosmic string spacetime. In order to do that we shall consider 
the non-relativistic limit of the modified Dirac equation in this spacetime. 
Two different situations will be considered: $(i)$ For $p>\ 1$, 
the electric current will be attracted to the cosmic string (disclination), 
while the linear charge distribution will be repelled by it. As the final 
configuration we expected that the current will be superposed to the cosmic
string. Taking $r_0=0$ in (\ref{A}) and its rotational, we obtain an
azimuthal magnetic field
\begin{equation}
\label{B}
\vec{B}(\vec{r})=\frac{2j_0\ p}{cr}\hat{\phi} \ .
\end{equation}
$(ii)$ For the case where $p<\ 1$, the linear charge distribution will be
attracted to the disclination and the current repelled. In this case we 
expected as final configuration to have the linear charge superposed to
the linear defect. The electric field produced by this configuration can also
be obtained by taking the negative gradient of the scalar potential 
considering $r_0=0$. This field is radial as shown below:
\begin{equation}
\label{E1}
\vec{E}=\frac{2\lambda \ p}r\hat{r} \ .
\end{equation}  

The effect of an arbitrary electromagnetic field in the quantum motion of a
neutral particle with anomalous magnetic and electric moments has been analysed 
by Anandan \cite{Anadan}. There, using the non-relativistic limit of the Dirac
equation Anandan was able to shown that the two-component wave-function associated
with the particle acquires an Aharonov-Bohm phase. The induced gauge field
presents a non-Abelian character which makes the analysis of the quantum and
classical motion, and the phase shift richer.
More recently Audretsch and Skarzhinsky \cite{Audretsch} have analysed the 
Aharonov-Bohm scattering of a charged particles and neutral atoms
with induced electric electric dipole moments in absorbing medium using the
Schrodinger equation in the $z=0$ plane. The analysis of a charged particle
which presents anomalous magnetic moment in the presence of three different
configurations of magnetic fields along the $z-$direction, has been developed
in the context of the non-relativistic Schrodinger equation \cite{Bordag},
and also relativistic Dirac equation \cite{Bordag1} by Bordag and Voropaev.
There they shown that bound states are present if the magnetic moment is 
anti-parallel to the magnetic fields. 

This paper is organized as follows: In Sect. $2$ we develop the formalism 
needed to analyse the problem under consideration, i.e., the non-relativistic
quantum motion of uncharged particle with anomalous magnetic moment in the
presence of magnetic and electric fields in a conical spacetime. Some 
mathematical details will be introduced in order to derive the non-relativistic 
limit of the modified Dirac equation as a Schrodinger-Pauli one, to analyse  
two different physical systems. We shall see that the equation is a
generalization of the similar one obtained by Anandan in a conical spacetime.
In Sect. $3$ we explicitly analyse the quantum motion of an uncharged two-component 
spinor field in azimuthal and radial, magnetic and electric fields, respectively. 
Because of the presence of the anomalous moment, we obtain for the first 
case, two coupled second order differential equations. Although looking simple 
this set of differential equations, we were not able obtain an exact solution. 
However, for large total angular quantum number, approximated solutions can
be provided. Moreover, for bound states solutions we could obtain the respective
quantized energy. For the second case, considering only electric field, we do
not find bound states, only scattering ones take place and a phase shift
can be provided. As we shall see, for this case the scattering amplitude associated 
with both components of the spinor field depend on the parameter $\alpha$, which
codifies the conical structure of the spacetime, and also on the anomalous
coupling factor $\kappa$. We leave for Sect. $4$ our most important remarks 
about our analysis of the system. In the Appendix we present some specific 
calculations used to obtain the non-relativistic limit of the modified Dirac 
equation in a conical spacetime in a more familiar and workable way.

\section{The Equation of Motion}

The electromagnetic interactions of mesons and nucleons depend strongly on 
their anomalous electric and magnetic moments. For specific case of the neutron 
its magnetic moment is $-1.91\mu_B$, whereas the free Dirac equation predicts
a zero value. It is possible to include the effect of such interaction in the
analysis of quantum motion if we abandon the principle of minimal coupling
\cite{BD}. For a flat spacetime instead of consider only the minimal 
electromagnetic coupling, it is added an extra dipole interaction term:
\begin{equation}
\label{mu}
\D=\gamma^{(a)}(i\hbar\partial_{(a)}-(e/c)A_{(a)}) \ \ \longrightarrow \ \ \D-
\kappa\frac{\mu_B}{2\hbar c}
\sigma^{(a)(b)}F_{(a)(b)} \ ,
\end{equation}
where $\kappa$ is the anomalous factor, $\sigma^{(a)(b)}=1/2[\gamma^{(a)}, 
\gamma^{(b)}]$, $\mu_B=e\hbar/2mc$ the Bohr magneton and $\gamma^{(a)}$ the 
Dirac matrices.

However for a non-flat spacetime, we have to include the spin connection in the
differential operator and define the respective Dirac matrices in this 
manifold. Taking a basis tetrad $e^\mu_{(a)}$ satisfying the condition
\begin{equation}
e^\mu_{(a)} e^\nu_{(b)}\eta^{(a)(b)}=g^{\mu\nu} \ ,
\end{equation}
the spin connection $\Gamma_\mu$ can be expressed by
\begin{equation}
\label{gamma}
\Gamma_\mu=-\frac14\gamma^{(a)}\gamma^{(b)}e^\nu_{(a)}e_{(b)\nu;\mu} \ ,
\end{equation}
and the $\gamma$ matrices in the curved spacetime by $\gamma^\mu=e^\mu_{(a)}(x)
\gamma^{(a)}$. In this way the modified Dirac equation in the curved space 
reads:
\begin{equation}
\left[i\hbar\gamma^\mu(\partial_\mu+\Gamma_\mu)-\frac ec\gamma^\mu A_\mu-
\kappa\frac{\mu_B}{2c}\sigma^{\mu\nu}F_{\mu\nu}-mc\right]\Psi(x)=0 \ .
\end{equation}
Because we have the intention to analyse the non-relativistic limit of this
Dirac equation {\it a la} procedure presented in \cite{BD}, let us first
to express it in terms of an Hamiltonian formalism. After some intermediate
steps we get:
\begin{equation}
\hat{H}\Psi(x)=i\hbar\partial_t\Psi(x) \ ,
\end{equation}
with
\begin{equation}
\label{H}
\hat{H}=c\alpha^i(-i\hbar\partial_i+\frac ecA_i)+eA_0-ic\hbar\gamma^0
\gamma^\mu\Gamma_\mu+\kappa\frac{\mu_B}2\gamma^0\sigma^{\mu\nu}F_{\mu\nu}
+\beta mc^2 \ ,
\end{equation}
where we have assumed that the time-time component of the metric tensor is
equal to $-1$, and defined $\alpha^i=\gamma^0\gamma^i$ and $\beta=\gamma^0$ 
as usual. The $\gamma-$matrices obey the anticommutator relations 
$\{\gamma^\mu, \ \gamma^\nu\}=-2g^{\mu\nu}$. 

Now after this brief introduction let us specialize in the cosmic string
spacetime whose the metric tensor components are given by (\ref{CS}). Adopting
the following representation to the $\gamma-$matrices in a flat spacetime
\begin{equation}
\gamma^{(0)}=\left( 
\begin{array}{cc}
1&0 \\
0&-1 
\end{array} \right) \ , \ \ 
\gamma^{(a)}=\left( 
\begin{array}{cc}
0&\sigma^{(a)} \\
-\sigma^{(a)}&0 
\end{array} \right) \ ,
\end{equation}
with $\sigma^{(a)}$ being the standard Pauli matrices and a particular basis
tetrad 
\begin{equation}
e^\mu_{(a)} = \left( \begin{array}{cccc}
  1 & 0 & 0 & 0 \\
  0 &\cos\phi &-\sin\phi/\alpha r & 0 \\ 
0 & \sin\phi & \cos\phi/\alpha r & 0 \\ 
0 & 0 & 0 & 1 
                      \end{array}
               \right) \ .
\label{Tetrada}
\end{equation}
For the cosmic string spacetime
\begin{equation}
\gamma^\mu\Gamma_\mu=-\frac{(\alpha^{-1}-1)}{2r}\gamma^r \ ,
\end{equation}
with
\begin{equation}
\gamma^r=\cos\phi\gamma^{(1)}+\sin\phi\gamma^{(2)}=\left(\begin{array}{cc}
0&\sigma^r \\
-\sigma^r&0 \end{array} \right) \ .
\end{equation}
Moreover the $\alpha-$matrices are now written as
\begin{equation}
\alpha^i=e^i_{(a)}\left(\begin{array}{cc}
0&\sigma^{(a)} \\
\sigma^{(a)}&0 \end{array} \right)=\left(\begin{array}{cc}
0&\sigma^i \\
\sigma^i&0 \end{array} \right) \ ,
\end{equation}
where $\sigma^i=(\sigma^r, \ \sigma^\phi, \ \sigma^z)$ are the Pauli matrices
in cylindrical coordinates obtained from the basis tetrad (\ref{Tetrada}).

Now let us study the non-relativistic limit of the modified Dirac equation. In
this limit we have to extract from the wave-function its temporal behavior 
consequence of the rest energy. So writing
\begin{equation}
\Psi=\ e^{-i(mc^2/\hbar t)}\left(\begin{array}{cc} \varphi \\
\chi\end{array}\right) \ ,
\end{equation}
we obtain the following differential equation:
\begin{eqnarray}
c\sigma^i\pi^i\left(\begin{array}{cc}\chi \\\varphi\end{array}\right)
+i\hbar\frac{(\alpha^{-1}-1)}{2r}c\sigma^r\left(\begin{array}{cc} \chi \\
\varphi\end{array}\right)&+&eA_0\left(\begin{array}{cc} \varphi \\
\chi\end{array}\right)+i\kappa\mu_B\frac{d\Phi(r)}{dr}\sigma^r\times\nonumber\\
\left(\begin{array}{cc} \chi \\-\varphi\end{array}\right)
-\kappa\mu_BF_{13}\alpha r\sigma^\phi\left(\begin{array}{cc} \varphi \\
-\chi\end{array}\right)&-&2mc^2\left(\begin{array}{cc} 0 \\
\chi\end{array}\right)=i\hbar\partial_t\left(\begin{array}{cc} \varphi \\
\chi\end{array}\right) \ ,
\end{eqnarray}
where $\pi^i=-i\hbar\partial_i+(e/c)A_i$. The ``small'' component of the
wave function, $\chi$, can be related with its ``large'' one, $\varphi$, by
\begin{equation}
\label{chi}
\chi=\frac1{2mc}\sigma^i\left[-i\hbar(\partial_i+\Gamma_i)+(e/c)A_i
+i\kappa(\mu_B/c)E_i\right]\varphi \ .
\end{equation} 
It is worth to mention that the approach developed here allows to consider the
two different physical configurations at the same time. For the case where
the magnetic interaction will be considered we discard the electric field in
our equations, and the opposite procedure if we are considering electric 
interaction. Moreover, because vector potential posses only component along the
$z-$direction and the electric field is radial, and because $g_{11}=g_{33}=
1$, we have $A^z=A_z$ and $E^r=E_r$, so we can see that $\sigma^rE^r=
\sigma^iE_i$. We are writing the general formalism above in order to be
complete, however, later we shall consider uncharged particle only, and these
observations will not be essential anymore.

The remaining equation is:
\begin{eqnarray}
c\sigma^i\left[-i\hbar(\partial_i+\Gamma_i)+(e/c)A_i-i\kappa(\mu_B/c)E_i
\right]\chi\nonumber\\
+eA_0\varphi-\mu_B\kappa F_{13}\alpha r\sigma^\phi\varphi=
i\hbar\partial_t\varphi \ .
\end{eqnarray}
Now substituting (\ref{chi}) into the above equation we obtain:
\begin{equation}
\hat{H}_{NR}\varphi=i\hbar\partial_t\varphi
\end{equation}
where
\begin{eqnarray}
\label{H0}
\hat{H}_{NR}&=&\frac{\sigma^i(\Pi_i-i\kappa(\mu_B/c)E_i)\sigma^j(\Pi_j+
i\kappa(\mu_B/c)E_j)}{2m}\nonumber\\
&+&eA_0-\mu_B\kappa\alpha rF_{13}\sigma^\phi ,
\end{eqnarray}
where now $\Pi_i=-i\hbar(\partial_i+\Gamma_i)+(e/c)A_i$. In the Appendix we
present specific calculations used to write the above Hamiltonian in a more
familiar and workable way. Our final expression to it is:
\begin{eqnarray}
\label{H}
\hat{H}_{NR}&=&\frac1{2m}\frac1{\sqrt {g^{(3)}}}\left(\Pi_i-
\kappa\frac{\mu_B}c(\vec{E}\times\vec{\sigma})_i\right)\left[g^{ij}
\sqrt{g^{(3)}}\left(\Pi_j-\kappa\frac{\mu_B}c(\vec{E}\times
\vec{\sigma})_j\right)\right]\nonumber\\
&+&\frac{\kappa\mu_B\hbar}{2mc}({\vec{\nabla}}\cdot{\vec{E}})
+\frac{\hbar e}{2mc}\alpha r\vec{B}\cdot\vec{\sigma}-eA^0
+\mu_B\kappa\alpha r\vec{B}\cdot\vec{\sigma}-\frac{\kappa^2\mu_B^2}{2mc^2}
\vec{E}^2 \ ,
\end{eqnarray}
where $g^{(3)}$ is the determinant of the spatial component of (\ref{CS}).

Having found the general expression to the non-relativistic Hamiltonian which
governs the motion of a charged particle with anomalous moment in the conical
spacetime, and in presence of a cylindrically symmetric radial and azimuthal
electric and magnetic fields, respectively, let us specialize in two
distinct situations already mentioned. So the first steps is to set
the electric charge equal to zero. The second one in fact involves two different
approaches which will be considered independently in the next section.

\section{The Non-relativistic Motion}

\subsection{Motion in a Azimuthal Magnetic Field}

This situation happens when the linear uniform electric current is running along
the $z-$axis in the cosmic string spacetime. For this case (\ref{H}) reads:
\begin{equation}
\hat{H}_{NR}=-\frac{\hbar^2}{2m}\frac1{\sqrt{g^{(3)}}}\nabla_i\left(g^{ij}
\sqrt{g^{(3)}}\nabla_j\right)+\mu_B\kappa\vec{B}\cdot\vec{\sigma} \ ,
\end{equation}
where $\nabla_i=\partial_i+\Gamma_i$. For the specific basis tetrad used 
here, the only non-vanishing spin connection is:
\begin{equation}
\Gamma_\varphi=i\frac{(1-\alpha)}2\sigma^z \ .
\end{equation}

As it has been pointed out by Anandan in \cite{Anadan}, here we also
can interpret the above Hamiltonian as the operator which governs the
motion of a ''charged'' particle coupled with $SU(2)$ gauge potential
$A_\mu=A_\mu^a\sigma^a$ in the conical spacetime. Comparing the above 
expressions with the ordinary definition to the operator
$\pi^i=-i\hbar\partial_i+q/cA_i$, and the electrostatic interaction $-qA_0$, 
we can identify the ``gauge potential'' as:
\begin{equation}
A_0^a=\mu_B(\kappa/q)B^a
\end{equation}
and 
\begin{equation}
A_i^a=\hbar c(1-\alpha)/2q\ \delta_i^\phi\ \delta_3^a \ ,
\end{equation} 
where the latter has purely geometric character. The two-component iso-spinor 
reads:
\begin{equation}
\phi=\left(\begin{array}{cc}
\phi_+ \\ \phi_- \end{array} \right) \ .
\end{equation}
After this identification, we pass now to develop the above Hamiltonian, which can 
be written as:
\begin{equation}
\hat{H}_{NR}=-\frac{\hbar^2}{2m}\nabla^2_{(\alpha)}-\frac{i\hbar}{2m}
\frac{(1-\alpha)}{\alpha^2r^2}\sigma^z\partial_\phi+\frac{\hbar^2}{2m}
\frac{(1-\alpha)^2}{4\alpha^2r^2}-\frac{2\mu_B\kappa j_0}c\sigma^\phi \ ,
\end{equation}
where $\nabla^2_{(\alpha)}=\partial_r^2+1/r\partial_r+(1/\alpha^2r^2)
\partial_\phi^2+\partial_z^2$, is the Laplace-Beltrami operator in the conical
space. In the above expression we have substituted the explicit expression to 
the magnetic field: Eq. (\ref{B}). Because this system presents a invariance under 
boosts along the $z-$direction the stationary states can be written as 
$\Psi(t,\vec{x})=e^{-iEt/\hbar}e^{ik_zz}u(r,\theta)$. Substituting this 
wave-function into the Schrodinger-Pauli equation we get the self-energy equation:
\begin{equation}
\label{Heff}
\hat{H}_{Eff}u(r,\theta)=Eu(r,\theta) \ ,
\end{equation}
with
\begin{eqnarray}
\label{Heff1}
\hat{H}_{Eff}&=&-\frac{\hbar^2}{2m}\left[\partial_r^2+\frac1r\partial_r+
\frac1{\alpha^2r^2}\partial_\phi^2\right]+\frac{\hbar^2k_z^2}{2m}
+\frac{i\hbar^2}{2m}\frac{(1-\alpha)}{\alpha^2r^2}\sigma^z\partial_\phi
\nonumber\\
&+&\frac{\hbar^2}{2m}\frac{(1-\alpha)^2}{4\alpha^2r^2}-\frac{2\mu_B\kappa j_0}
c\sigma^\phi \ .
\end{eqnarray}

For this system a total angular momentum operator can be defined as the 
sum of the orbital and iso-spin parts 
\begin{equation}
\label{J}
\hat{J}=-i\partial_\theta+(1/2)\sigma^z \ .
\end{equation}
It can be  show that $\hat{J}$ commutes with the effective Hamiltonian. So it is 
possible to express the eigenfunctions of the effective two dimensional 
Hamiltonian, $\hat{H}_{Eff}$, in terms of the eigenfunction of $\hat{J}$. These 
functions have the form
\begin{equation}
\label{u}
u_j(r,\theta)=\frac{e^{ij\theta}}{\sqrt r}\left(\begin{array}{cc}
u_+(r)\ e^{-i\theta/2} \\ iu_-(r)\ e^{i\theta/2} \end{array} \right) \ ,
\end{equation}
with $j=n+1/2=\ \pm 1/2,\ \pm 3/2, \ ...\ $. 

Substituting (\ref{Heff1}) and (\ref{u}) into (\ref{Heff}) we obtain, after a 
few steps, a set of two coupled radial differential equations:
\begin{equation}
u_+''(r)-\frac{j(j-\alpha)}{\alpha^2r^2}u_+(r)-\frac{4m\mu_B\kappa j_0}
{\alpha\hbar^2cr}u_-(r)+k^2u_+(r)=\ 0 
\end{equation}
and
\begin{equation}
u_-''(r)-\frac{j(j+\alpha)}{\alpha^2r^2}u_-(r)-\frac{4m\mu_B\kappa j_0}
{\alpha\hbar^2cr}u_+(r)+k^2u_-(r)=\ 0 \ ,
\end{equation}
where $k^2=2m{\cal{E}}/\hbar^2$, being ${\cal{E}}=E-\hbar^2k^2/2m$. Setting
$\kappa=0$ we obtain two non-coupled differential equations. Moreover, this
system is symmetric under the transformation $j \to -j$ and $u_+ 
\rightleftharpoons\ u_-$. The asymptotic behavior for both solutions can be 
obtained by taking $r \to\ \infty$. For both equations we get
\begin{equation}
u_\pm''+k^2u_\pm=\ 0 \ ,
\end{equation}
which gives rise to possibilities: $(i)$ scattering states, $u_\pm\ \simeq\ 
e^{\pm ikr}$ and $(ii)$ bound states if $k^2=-k^2_B$ with $u_\pm\ \simeq\ 
e^{-k_Br}$. It is of our interest in this sub-section to analyse the possibility 
of this system to present bound states solutions. In this way, defining a 
dimensionless variable $z=k_Br$, we can in rewrite the two differential equations 
above as:
\begin{equation}
\frac{d^2u_+}{dz^2}-\frac{j(j-\alpha)}{\alpha^2z^2}u_+-\frac\delta zu_-
-u_+=0 
\end{equation}
and
\begin{equation}
\frac{d^2u_-}{dz^2}-\frac{j(j+\alpha)}{\alpha^2z^2}u_--\frac\delta zu_+
-u_-=0 \ ,
\end{equation}
with
\begin{equation}
\label{delta} 
\delta=\frac{4m\mu_B\kappa j_0}{\alpha\hbar^2k_Bc} \ .
\end{equation}

Defining 
\begin{equation}
\label{Psi}
\Psi(r)=\left(\begin{array}{cc}
u_+(r) \\ u_-(r) \end{array} \right) \ ,
\end{equation}
we can write these two equations above in terms of a $2\times 2$ Hermitian 
Hamiltonian operator:
\begin{equation}
\hat{H}\Psi(r)=-\Psi(r) \ ,
\end{equation}
with
\begin{equation}
\hat{H}=-I_{(2)}\left(\frac{d^2}{dz^2}-\frac{j^2}{\alpha^2z^2}\right)-
\sigma^{(3)}\frac{j}{\alpha z^2}+\sigma^{(1)}\frac{\delta}{z} \ ,
\end{equation}
being $I_{(2)}$ the identity matrix. Notice that the above system resemble 
with the Schrodinger equation for a particle with mass $M=1/2$ and $E=-1$ in 
the presence of a non-Abelian Coulomb potential. This potential has strength 
$\delta$ and couples both components of the effective radial wave-function 
(\ref{Psi}). Considering this analogy, the Coulomb potential will be attractive 
for negative value of $\delta$ and repulsive in the opposite case. Because we are
interested to analyse the possibility of this system to present bound states,
let us assume $\kappa$ negative. This is what happens with the neutron, which
presents $\kappa=-1.91$. Taking $\delta=-|\delta|$ in previous equations we get:
\begin{equation}
\frac{d^2u_+}{dz^2}-\frac{j(j-\alpha)}{\alpha^2z^2}u_++\frac{|\delta|} zu_-
-u_+=0 
\end{equation}
and
\begin{equation}
\frac{d^2u_-}{dz^2}-\frac{j(j+\alpha)}{\alpha^2z^2}u_-+\frac{|\delta|} zu_+
-u_-=0 \ .
\end{equation}

Although looks simple this set of coupled differential equation, we could 
not obtain an exact solution for it. However our main interest
in this paper is to investigate the possibility of this system to present
bound states. In this sense we can proceed on this analysis to the case
where the total angular quantum number $|j|$ is much grater than $\alpha$. 
Accepting this situation, two distinct classes of solutions can be analyzed:
$i)$ $u_+=-u_-$ and $ii)$ $u_+=u_-$. For the first class the effective
Coulomb potential becomes repulsive in both differential equations, so no
bound state can be formed. However for the second class, it becomes attractive
making possible the existence of bound states. For the latter both equations can
be written as:
\begin{equation}
\frac{d^2 f(z)}{dz^2}-\frac{j^2}{\alpha^2 z^2}f(z)+\frac{|\delta|}z
f(z)-f(z)=0 \ .
\end{equation}
The solution to this equation regular at origin is:
\begin{equation}
f(z)=M_{|\delta|/2,\sqrt{1/4+j^2/\alpha^2}}(2z) \ ,
\end{equation}
where $M_{\lambda,\mu}(z)$ is the Whittaker function \cite{Gradshteyn}, which
is expressed in terms of confluent hypergeometric function as:
\begin{equation}
M_{\lambda,\mu}(z)=z^{\mu+1/2}e^{-z/2}\Phi(\mu-\lambda+1/2, \ 2\mu+1, \ z) \ .
\end{equation}
In order to obtain bound states solution, one has to impose specific condition
on the parameter $\mu-\lambda+1/2$ to terminate the series, transforming it 
in a polynomial. This condition is to make this parameter equal to a non
positive integer number. For our case this condition provides
\begin{equation}
\sqrt{\frac14+\frac{j^2}{\alpha^2}}+\frac12-\frac{|\delta|}2\approx \ \frac{|j|}
\alpha+\frac12-\frac{|\delta|}2= \ - n \ ,
\end{equation}
$n$ being $0, \ 1, \ 2, \ ...$. Substituting the expression to $\delta$
given (\ref{delta}), $k_B=\sqrt{2m|{\cal{E}}|}/\hbar$ and after some intermediate
steps we find:
\begin{equation}
\label{E}
{\cal{E}}=E-\frac{\hbar^2k_z^2}{2m}=-\frac{8m\mu_B^2\kappa^2j_0^2}
{\alpha^2\hbar^2c^2(2n+1+2|j|/\alpha)^2} \ .
\end{equation}

\subsection{Motion in the Radial Electric Field}

This situation happens in disclinated medium with $\alpha$ bigger than
unity. In this case a linear electric charge distribution may be localized
along the $z-$axis. For this case (\ref{H}) becomes:
\begin{eqnarray}
\label{H00}
\hat{H}_{NR}&=&-\frac{\hbar^2}{2m}\frac1{\sqrt{g^{(3)}}}\left(\nabla_i
-i\frac{A_i}{\hbar}\right)\left[g^{ij}\sqrt{g^{(3)}}\left(\nabla_j
-i\frac{A_j}{\hbar}\right)\right]\nonumber\\
&-&\frac{\kappa^2\mu_B^2}{2mc^2}\vec{E}^2 \ ,
\end{eqnarray}
which can be seen as a Hamiltonian that governs the motion of a $SU(2)$ 
''charged'' iso-spinor particle interacting with a non-Abelian vector
potential 
\begin{equation}
A_i=c\hbar\frac{(1-\alpha)}{2q}\sigma^z\delta_i^\phi-\frac{a_ic}q
\end{equation}
with 
\begin{equation}
a_1=0\ ,  \ \ a_2=-\frac{\kappa\mu_B\alpha r}cE^r\sigma^z \ , \ \
a_3=\frac{\kappa\mu_B\alpha r}cE^r\sigma^\phi 
\end{equation}
and Abelian scalar potential
\begin{equation}
A_0=-\frac{\kappa^2\mu_B^2}{2mc^2q}{\vec{E}}^2 \ 
\end{equation}
in conical spacetime. $q$ being the ''electric'' charge associated with these 
potentials. In (\ref{H00}) we have dropped the term ${\vec{\nabla}}\cdot{\vec{E}}$, 
which is proportional to the electric charge density of the source
giving by a delta-function concentrated on the defect: ${\vec{\nabla}}\cdot
{\vec{E}}=2(\lambda/r)\delta(r)$. This procedure is allowed for wave-function
that vanish at origin. However, as will be mentioned later, by constructing
adjoint extension of the above Hamiltonian, specific irregular solutions are 
accepted.

The complete expansion to the above Hamiltonian contains many terms which can
be conveniently grouped. Inside the brackets there appears a term exactly twice the
similar one square in the electric field however with positive sign. So, after
some intermediate steps the Hamiltonian can be written as
\begin{eqnarray}
\hat{H}_{NR}&=&-\frac{\hbar^2}{2m}\nabla^2_{(\alpha)}-\frac{i\hbar^2(1-\alpha)}
{2m}\frac{\sigma^z}{\alpha^2r^2}\partial_\phi+\frac{\hbar^2}{2m}
\frac{(1-\alpha)^2}{4\alpha^2r^2}
-\frac{i\hbar\kappa\mu_BE^r\sigma^z}{mc\alpha r}\partial_\phi\nonumber\\
&+&\frac{\hbar\kappa\mu_BE^r(1-\alpha)}{2mc\alpha r}+
\frac{i\hbar\kappa\mu_B\alpha rE^r\sigma^\phi}{mc}\partial_z
+\frac{\kappa^2\mu_B^2}
{2mc^2}(E^r)^2 \ .
\end{eqnarray}

As in the previous analysis this system presents a invariance under boosts 
along the $z-$direction, so its stationary states can be written as $\Psi(t,\vec{x})=
e^{-iEt/\hbar}e^{ik_zz}u(r,\theta)$. Substituting this wave-function into the 
Schrodinger-Pauli equation 
\begin{equation}
\hat{H}_{Eff}u(r,\theta)=Eu(r,\theta) \ ,
\end{equation}
we obtain
\begin{eqnarray}
\hat{H}_{Eff}&=&-\frac{\hbar^2}{2m}\left[\partial_r^2+\frac1r\partial_r+\frac1
{\alpha^2r^2}\partial_\phi^2\right]-\frac{i\hbar^2(1-\alpha)}{2m}\frac{\sigma^z}
{\alpha^2r^2}\partial_\theta+\frac{\hbar^2}{2m}\frac{(1-\alpha)^2}{4\alpha^2r^2}
\nonumber\\
&-&\frac{i\hbar\kappa\mu_BE^r\sigma^z}{mc\alpha r}\partial_\phi
+\frac{\hbar\kappa\mu_BE^r(1-\alpha)}{2mc\alpha r}-\frac{\hbar\kappa
\mu_B\alpha rE^r\sigma^\phi k_z}{mc}+\frac{\kappa^2\mu_B^2}{2mc^2}(E^r)^2 \ .
\end{eqnarray}

As in the previous analysis we can verify that the total angular operator 
$\hat{J}$ commutes with $\hat{H}_{Eff}$ above, consequently we can use the 
same procedure expressing its eigenfunctions as exhibited in (\ref{u}). 
Moreover, from the explicit form of the effective Hamiltonian, 
we can observe that choosing $k_z=0$, it becomes diagonal, consequently the
upper and lower components of the wave-function do not interact among themselves. 
So, for simplicity only, let us consider this possibility. Finally taking into 
account all the observations mentioned, and after some intermediate steps 
we arrive at:
\begin{equation}
u_+''(r)-\frac{j(j-\alpha)}{\alpha^2r^2}u_+(r)-\frac{\kappa\mu_B(2j-\alpha)}
{\hbar c\alpha r}E^r u_+(r)-\frac{\kappa^2\mu_B^2(E^r)^2}{\hbar^2c^2}u_+(r)
=- k^2u_+(r) 
\end{equation}
and
\begin{equation}
u_-''(r)-\frac{j(j+\alpha)}{\alpha^2r^2}u_-(r)+\frac{\kappa\mu_B(2j+\alpha)}
{\hbar c\alpha r}E^r u_-(r)-\frac{\kappa^2\mu_B^2(E^r)^2}{\hbar^2c^2}u_-(r)
=- k^2u_-(r) 
\end{equation}
with $k^2=2mE/\hbar^2$. Substituting the expression to the electric field 
(\ref{E1}) and defining a new variable $z=kr$ we can notice that both 
equations take the form
\begin{equation}
\frac{d^2u_\pm(z)}{dz^2}-\frac{\gamma_\pm}{z^2}u_\pm(z)+u_\pm(z)= \ 0 \ ,
\end{equation}
where
\begin{equation}
\gamma_\pm=\frac{j(j\mp\alpha)}{\alpha^2}\pm\frac{2\kappa\mu_B\lambda(2j\mp\alpha)}
{\alpha^2\hbar c}+\frac{4\kappa^2\mu_B^2\lambda^2}{\alpha^2\hbar c^2} \ .
\end{equation}
The general solution for these equations are given in terms of Bessel functions:
$u_\pm(z)=\sqrt{z}(A_{\nu_\pm}J_{\nu_\pm}(z)+B_{\nu_\pm}J_{-\nu_\pm}(z))$ with
\begin{equation}
\nu_\pm=\sqrt{\frac14+\gamma_\pm}=
\left|\frac n\alpha+\frac{1\mp\alpha}{2\alpha}\pm\frac{2\kappa\mu_B\lambda}
{\alpha\hbar c}\right| \ ,
\end{equation}
where we have substituted $j=n+1/2$.

The scattering amplitude associated with this system can be evaluated by using
partial-wave analysis. A few years ago, Hagen showed that, similarly with happens 
in the Aharonov-Bohm scattering for spin$-1/2$ particles by an idealized solenoid 
\cite{H1}, the admissible solutions to the above equations are the regular one, 
unless $\nu_\pm$ be smaller than unity \cite{H2}. In this case, irregular solutions 
becomes the only ones acceptable. So, to calculate the scattering amplitude 
associated with this system by using partial-wave analysis, both kind of solutions 
must be taken into account. Defining $2\kappa\mu_B\lambda/(\hbar c)=N+\delta$, 
with $N$ being an integer number and $0\leq \delta<1$, the acceptable singular 
solutions are given by $n=-N-1$ to $u_+(z)$ and by $n=N$ to $u_-(z)$. For both, 
the order of the singular Bessel function is
\begin{equation}
\nu=-\frac{1+\alpha}{2\alpha}+\frac\delta\alpha=-\beta \ ,
\end{equation}
provided $0<\ \beta<\ 1$ \footnote{The system of spin$-1/2$ particle with anomalous
magnetic moment in presence of a idealized solenoid has been analyzed in \cite{VB}
in a non-relativistic context. There by applying the method of von Neumann to 
provide a self-adjoint extension to the Hamiltonian, they also shown that singular 
solutions to the wave-equation are accepted.}.  

The regular effective two dimensional wave function (\ref{u}) is now written as:
\begin{equation}
\label{uj}
u_j(r,\theta)=u_n(r,\theta)=e^{in\theta}\left(\begin{array}{cc}
c^n_+J_{\nu_+}(kr) \\ c^n_-J_{\nu_-}(kr)\ e^{i\theta} 
\end{array} \right) \ ,
\end{equation}
where the coefficients $c_n^+$ and $c_n^-$ can be determined by imposing that the
complete wave-function given below, for large values of $r$, behaves as an incident 
plane wave coming from the right:
\begin{equation}
\label{psi}
\psi(r,\theta)={\sum}' u_n(r,\theta)+\left(\begin{array}{cc}
e^{-i(N+1)\theta}e^{i(\pi/2)\beta}J_{-\beta}(kr) \\ 
e^{i(N+1)\theta}e^{i(\pi/2)\beta}J_{-\beta}(kr)\end{array} \right) \ .
\end{equation}
In the above expression the prime on the summation indicates the exclusion of the 
irregular solutions. Using the asymptotic behavior for the Bessel functions, the 
coefficients $c_n^+$ and $c_n^-$ are given by:
\begin{eqnarray}
\label{c}
c_n^+&=&e^{-i(\pi/2)\nu_+}\nonumber\\
c_n^-&=&e^{-i(\pi/2)\nu_-}
\end{eqnarray}
with $\theta=\pi+\theta'$.

In order to calculate the complete scattering amplitude associated with this
system, it is more convenient to write (\ref{psi}) as shown below:
\begin{equation}
\label{psi1}
\psi(r,\theta)=\ \psi_R(r,\theta)+\ \psi_I(r,\theta) \ ,
\end{equation}
with
\begin{equation}
\label{psiR}
\psi_R(r,\theta)=\sum_{n=-\infty}^{\infty} u_n(r,\theta)
\end{equation}
and
\begin{equation}
\label{psiI}
\psi_I(r,\theta)=\left(e^{i(\pi/2)\beta}J_{-\beta}(kr)-e^{-i(\pi/2)\beta}J_\beta(kr)
\right)\left(\begin{array}{cc}
e^{-i(N+1)\theta} \\ e^{i(N+1)\theta}\end{array} \right) \ .
\end{equation}
So to obtain the complete scattering amplitude, two distinct contributions can
be separately calculated. 

The scattering amplitude associated with the regular solutions, $f_R(\theta)$, 
can be calculated by using the phases shift $\delta_n^+$ and $\delta_n^-$, 
obtained from the asymptotic behavior of each component of (\ref{psiR}). It is:
\begin{equation}
\label{fR}
f_R(\theta)=\frac1{\sqrt{2\pi ik}}\sum_{n=-\infty}^\infty e^{in\theta}
\left(\begin{array}{cc}e^{2i\delta^+_n}-1 \\ e^{2i\delta^-_n}-1
\end{array} \right)=\left(\begin{array}{cc}f^+(\theta) \\ f^-(\theta)
\end{array} \right) \ .
\end{equation}
with
\begin{eqnarray}
\label{delta1}
\delta_n^+=-\frac\pi 2\nu_++\frac\pi 2|n|\nonumber\\
\delta_n^-=-\frac\pi 2\nu_-+\frac\pi 2|n| \ .
\end{eqnarray}
From the above expressions we can see that both components of the wave-function
have different phase shifts, and these phase shifts present two distinct
contributions: from the geometry of the spacetime itself codified by the
parameter $\alpha$, and by the anomalous factor $\kappa$. Due to its non-Abelian
character, the Aharonov-Bohm phase associated with the latter presents opposite 
sign. Because of the dynamics of each component does not depend on the other,
their respective scattering amplitude can be calculated independently. Let us now
develop these calculations using the partial-wave analysis:
\begin{eqnarray}
\delta_n^+=\left\{ 
\begin{array}{ll}
-n\frac\omega2-\frac{\upsilon^+}2 \ , \ n\geq \ - N \ , & \\
n\frac\omega2+\frac{\upsilon^+}2 \ , \ n< - \ N& \\ 
\end{array}
\right.
\end{eqnarray} 
and
\begin{eqnarray}
\delta_n^-=\left\{ 
\begin{array}{ll}
-n\frac\omega2+\frac{\upsilon^-}2 \ , \ n\geq \  N \ , & \\
n\frac\omega2-\frac{\upsilon^-}2 \ , \ n< \ N& \\ 
\end{array}
\right.
\end{eqnarray} 
where $\omega=\pi(\alpha^{-1}-1)$ and $\upsilon^\pm=N\pi/\alpha\pm\omega/2
+\pi\delta/\alpha$. Here we shall adopt the traditional approach to procedure 
the summation. We separate the complete sum in $n$ from $-N$ to $\infty$ and 
from $-\infty$ to $-N-1$ to the upper component, and from $N$ to $\infty$ and from 
$-\infty$ to $N-1$ to the lower one, using the respective value to the phase shift 
in the specific intervals; moreover in evaluating the sums we replace $\theta$ 
by $\theta+i\epsilon$ in the first summations and by $\theta-i\epsilon$ in the
second. All the summations can be done by usual way and after taken the limit 
$\epsilon\to \ 0$. Considering only scattering particle with $\theta\neq \ 0 \ , \ 
\omega$ and $-\omega$, we get:
\begin{equation}
f_R^+(\theta)=\frac{e^{-i(N+1/2)\theta}}{2i\sqrt{2\pi i k}}\left[
\frac{e^{i\pi\delta/\alpha-iN\pi}}{\sin(\frac{\theta+\omega}2)}-
\frac{e^{-i\pi\delta/\alpha+iN\pi}}{\sin(\frac{\theta-\omega}2)}\right]
\end{equation}
and
\begin{equation}
f_R^-(\theta)=\frac{e^{i(N-1/2)\theta}}{2i\sqrt{2\pi i k}}\left[
\frac{e^{i\pi\delta/\alpha+iN\pi}}{\sin(\frac{\theta-\omega}2)}-
\frac{e^{-i\pi\delta/\alpha-iN\pi}}{\sin(\frac{\theta+\omega}2)}\right]
\end{equation}

Considering the case with $\alpha=1$, $\omega=0$, we have
\begin{equation}
f_R^+(\theta)=(-1)^N\frac{e^{-i(N+1/2)\theta}}{\sqrt{2\pi i k}}\frac{\sin(\pi\delta)}
{\sin(\theta/2)} 
\end{equation}
and
\begin{equation}
f_R^-(\theta)=(-1)^N\frac{e^{i(N-1/2)\theta}}{\sqrt{2\pi i k}}\frac{\sin(\pi\delta)}
{\sin(\theta/2)} \ .
\end{equation}

The expression (\ref{fR}) to the scattering amplitude, has been obtained by 
standard procedure which consists to substitute the asymptotic behavior for the
Bessel function in (\ref{psiR}); however it has been pointed out by Hagen 
\cite{Hagen} that for large values of the order of the Bessel function, this 
procedure becomes inappropriate. One has to sum up first and then take the limit 
$kr\to\infty$. Using the integral representation to the Bessel function 
\cite{Gradshteyn}
\begin{equation}
\label{J}
J_\nu(z)=\frac1\pi\int_0^\pi\ dt\cos(\nu t-z\sin t)-\frac{\sin\nu t}\pi
\int_0^\infty\ dt e^{-\nu t-z\sinh t} \ ,
\end{equation}
this approach can be applied. Because we are interested to calculate the
scattering wave-function which behaves for large $r$ as $e^{ikr}/\sqrt{r}$, we
need only to use the integral $1/(2\pi)\int_0^\pi dte^{-i\nu t+iz\sin t}$.
Substituting this integral in (\ref{psiR}) and summing the geometric  
series according to the procedure indicate above, we obtain two integrals 
contributions for each component $f_R^+$ and $f_R^-$. Changing the variable $t$ 
to $t'+\pi/2$ and taking the limit $kr\to\infty$, both integrals can be 
approximately evaluated providing for the scattering amplitude the same result 
as obtained before. 

The contribution to the scattering amplitude due to (\ref{psiI}) can also be
calculated separately for both components. Their results are easily obtained:
\begin{equation}
f_I^+(\theta)=-(-1)^Ne^{-i(N+1)\theta}\sqrt{\frac{2i}{\pi k}}\sin(\pi\beta)
\end{equation}
and
\begin{equation}
f_I^-(\theta)=-(-1)^Ne^{i(N+1)\theta}\sqrt{\frac{2i}{\pi k}}\sin(\pi\beta) \ .
\end{equation}

\section{Concluding Remarks}

In this paper we have explicitly analysed the non-relativistic quantum motion
of a neutral particle possessing an anomalous magnetic moment in the presence of 
magnetic and electric fields in a conical spacetime. Two different physical 
systems have been investigated. The particle interacting with an azimuthal 
magnetic field and a radial electric one. In the first case we have shown that 
bound states between the neutral particle and uniform electric current may be 
formed, and the respective self-energy calculated. From our result, Eq. 
(\ref{E}), we observe that because $\alpha\neq \ 1$, the effective quantum number 
$2n+1+2|j|/\alpha$ is not an integer one. This fact reduces the degree of degeneracy 
of this ''Coulomb'' problem. In the second analysis we have shown that the 
electric interaction, together with the geometric one, contribute only to the 
scattering phase shift. Also from (\ref{delta1}) we can see that the scattering 
phase shifts associated with both components of the fermionic wave-function depend 
on the geometry itself by the parameter $\alpha$ and by the interaction between 
the anomalous magnetic moment and the electric field. Because of the non-Abelian 
character associated with the latter, the scattering amplitude for both components 
present opposite sign in their respective phases for regular functions. Another
point that we should mention is that Hagen has considered in his analysis about the
Aharonov-Casher scattering \cite{H2} two degrees of freedom associated with the
spin. Here we have explicitly use one. Including both degrees we would have 
reached the same conclusions as we did in sub-sections $(3.1)$ and $(3.2)$.

Bound states solutions for a neutral particle in a magnetic field of linear 
current density have been first investigated by Pron'ko and Stroganov \cite{Pronko}
in a flat spacetime. There, exploring a hidden $O(3)$ dynamical symmetry to the
system, the authors found an exact energy spectrum, which in our notation reads:
\begin{equation}
E_n=-\frac{2m\mu_B^2\kappa^2j_0^2}{\hbar^2c^2n^2} \ ,
\end{equation}
being $n=1, \ 2, \ 3$... . A few years latter Bl\"umel and Dietrich \cite{Blumel} 
returned to this system, and considering the limit $|j|>>1$ in the set of 
differential equations, they found an approximated solution to the energy spectrum 
similar to ours, without the factor $\alpha$. Moreover, using variational method 
they also showed that the improved results to the energy spectrum has no significant 
difference in the limit of large angular quantum number. Unfortunately, because we 
are considering this system in a conical spacetime, we could not find to the 
effective two-dimensional Hamiltonian operator (\ref{Heff1}) the same degree
of symmetry found in \cite{Pronko}. So only approximated results to the energy
spectrum has been provided. 

The main result of this analysis was to provide energy spectrum to a
neutral particle in a magnetic field of a linear current density superposed
to a conical singularity. This analysis  may be useful to study the quantum
motion of neutral spin$-1/2$ particle in disclinated medium in presence of 
electric current.

The analysis of Landau levels of charged particles in the presence of disclinations
has been developed a few years ago \cite{Furtado}. There bound states were
found, even considering the effect of a repulsive self-interaction. Here we
obtained bound states for neutral particle as a consequence of the interaction
between its anomalous magnetic moment with an azimuthal magnetic field. 

Finally we would like to finish this paper by saying that a similar analysis,
could also be develop if instead of an anomalous magnetic moment the particle
had an anomalous electric dipole moment. In this case we had to to change
$F_{(a)(b)}$ in (\ref{mu}) by its dual $^*F_{(a)(b)}=\frac12\epsilon_{(a)(b)
(c)(d)}F^{(c)(d)}$. 

\section*{Acknowledgment}

I would like to to thank Conselho Nacional de Desenvolvimento Cient\'{\i}fico
e
 Tecnol\'ogico (CNPq.) for partial financial support.

\appendix

\section{Development of the Non-relativistic Hamiltonian} 

Here we shall present some intermediate steps used to obtain (\ref{H}) from 
(\ref{H0}). First we need to find
\begin{equation}
\Pi_i\sigma^j=-i\hbar\nabla_i\sigma^j+\frac ec A_i\sigma^j \ ,
\end{equation}
with $\nabla_i=\partial_i+\Gamma_i$; however
\begin{eqnarray}
\nabla_i\sigma^j&=&\partial_i\sigma^j+\sigma^j\partial_i+\Gamma_i
\sigma^j\nonumber\\
&=&\sigma^j\nabla_i+\partial_i\sigma^j+[\Gamma_i, \ \sigma^j] \ .
\end{eqnarray}
So we need $\sigma^i(\partial_i\sigma^j)$. Writing $\sigma^j=e^i_{(a)}
\sigma^{(a)}$, being $\sigma^{(a)}$ the constant Pauli matrix and $e^i_{(a)}$
the basis tetrad. We find 
\begin{equation}
\label{A1}
\sigma^i(\partial_i\sigma^j)=\frac1{\alpha r}g^{ij}+i(1-\alpha)g^{2j}
\sigma^z \ .
\end{equation}
We also need to find the commutator $[\Gamma_i, \ \sigma^j]$, where for this
system the only non-vanishing spin connection is:
\begin{equation}
\Gamma_\phi=i(1-\alpha)\frac{\sigma^z}2  \ .
\end{equation}
So, $\sigma^i[\Gamma_i, \ \sigma^j]=i(1-\alpha)/2\sigma^\phi[\sigma^z, \ \sigma^j]$.
Again, expressing the $\sigma-$matrices in the conical spacetime in terms of the 
constants one, and developing the the commutator, we get: 
\begin{equation}
\label{A2}
\sigma^i[\Gamma_i,\ \sigma^j]=-\frac{1-\alpha}{\alpha r}g^{1j}-
i(1-\alpha)g^{2j}\sigma^z \ .
\end{equation}
Adding (\ref{A1}) and (\ref{A2}) we obtain
\begin{equation}
\sigma^i(\partial_i\sigma^j)+\sigma^i[\Gamma_i, \ \sigma^j]=\frac1rg^{1j} \ .
\end{equation}

After we have done this procedure, we have written (\ref{H0}) in terms of
$\sigma^i\sigma^j(\Pi_i-i\kappa\mu_B/cE_i)(\Pi_j+i\kappa\mu_B/cE_j)$.
Using the relation
\begin{equation}
\sigma^i\sigma^j=g^{ij}+i\frac{\epsilon^{ijm}}{\sqrt{g^{(3)}}}
g_{mk}\sigma^k \ ,
\end{equation}
being $g^{(3)}$ is the determinant of the space section of (\ref{CS}) and
$\epsilon^{ijm}$ the totally anti-symmetric symbol with $\epsilon^{123}=1$.
We could have stopped our calculation at this point; however, in order to
compare our expression with similar one obtained by Anandan in \cite{Anadan} to
ordinary flat spacetime, we decided to present our result in terms of the
modified linear operator $\Pi_i-\kappa(\mu_B/c)(\vec{E}\times\vec{\sigma})_i$.
So we can observe that (\ref{H}) is a generalization of the Eq. $(8)$ of
\cite{Anadan} written in a conical spacetime.


\begin{thebibliography}{99}
\bibitem{Vilenkin} A. Vilenkin, Phys. Rep. {\bf 121}, 263 (1985).
\bibitem{Kleiner} H. Kleinert, {\it Gauge Field in Condensed Matter} (World
Scientific, Singapore, 1989).
\bibitem{Kleman} M. Kleman, {\it Points, Lines and Walls} (Wiley, New York,
1983).
\bibitem{Katanaev} M. O. Katanaev and I. V. Volovich, Ann. Phys. (NY)
{\bf 216}, 1 (1992). 
\bibitem{Vilenkin1} A. Vilenkin, Phys. Rev. D {\bf 23}, 852 (1981); W. A. 
Hiscock, {\it ibid} {\bf 31}, 3288; B. Linet, Gen. Relativ. Gravit. {\bf 17},
1109 (1985).
\bibitem{Linet} B. Linet, Phys. Rev. D {\bf 33}, 1833 (1986).
\bibitem{Smith} A. G. Smith, in {\it Proceedings of Symposium on the Formation
and Evolution of Cosmic Strings}, ed. by G. W. Gibbons, S. W. Hawking and 
T. Vachaspati (Cambridge University Press, Cambridge, England, 1990).
\bibitem{Mello} E. R. Bezerra de Mello, V. B. Bezerra, C. Furtado and F.
Moraes, Phys. Rev. D {\bf 51}, 7140 (1995).
\bibitem{Dowker} J. S. Dowker, Nuovo Cimento B {\bf 52}, 129 (1967); L. H. 
Ford and A. Vilenkin, J. Phys. A {\bf 14}, 2353 (1981); V. B. Bezerra,
Phys. Rev. D {\bf 35}, 2031 (1987).
\bibitem{Frolov} V. P. Frolov and E. M. Serebryani, Phys. Rev. D {\bf 35},
1833 (1987); D. D. Harari and V. D. Skarzhinsky, Phys. Lett. B {\bf 240},
332 (1990).
\bibitem{Sokolov} D. D. Sokolov and A. A. Starobinsk, Sov. Phys. Dokl. 
{\bf 22}, 312 (1977).
\bibitem{Anadan} J. Anandan, Phys. Lett. A {\bf 138}, 347 (1989).
\bibitem{Audretsch} J. Audretsch and V. D. Skarzhinski, Phys. Rev. A {\bf 60},
1854 (1999).
\bibitem{Bordag} M. Bordag and S. Voropaev, J. Phys.: Math. Gen. {\bf 26}, 7637
(1993).
\bibitem{Bordag1}  M. Bordag and S. Voropaev, Phys. Lett. B {\bf 333}, 238 (1994). 
\bibitem{BD} J. D. Bjorken and S. D. Drell, {\it Relativistic Quantum 
Mechanics} (McGraw-Hill, New York, 1964). Pg. $241$.
\bibitem{Gradshteyn} I. S. Gradshteyn and I. M. Ryzhik, {\it Tables of
Integrals, Series, and Products}. (Academic Press, New York, (19080));
{\it Handbook of Mathematical Functions}, 9th ed. Edited by M.
Abramowitz and I. A. Stegum (Dove, New York, 1972).
\bibitem{H1} C. Hagen, Phys. Rev. Lett. {\bf 64}, 503 (1990).
\bibitem{H2} C. Hagen, Phys. Rev. Lett. {\bf 64}, 2347 (1990). 
\bibitem{VB} S. A. Voropaev and M. Bordag, Sov. Phys. JETP {\bf 78}(2),
127 (1994). 
\bibitem{Hagen} C. R. Hagen, Phys. Rev. D {\bf 41}, 2015 (1990).
\bibitem{Pronko} G. P. Pron'ko and Yu. G. Stroganov, Sov. Phys. JETP
{\bf 45}(6), 1075 (1977).
\bibitem{Blumel} R. Bl\"umel and K. Dietrich, Phys. Lett. A {\bf 139}, 236
(1989).
\bibitem{Furtado} C. Furtado, B. G. C. da Cunha, F. Moraes, E. R. Bezerra de
Mello and V. B. Bezerra, Phys. Lett. A {\bf 195}, 90 (1994).

\end{thebibliography}
\end{document}